\documentclass[twocolumn]{revtex4}

\usepackage{amsmath}
\usepackage{graphicx}
\usepackage{color}
\usepackage{float}
\usepackage{braket}
\usepackage{amssymb}
\usepackage{bbold}
\newcommand{\ua}{\uparrow}
\newcommand{\da}{\downarrow}

\DeclareMathOperator{\re}{Re}

\DeclareMathOperator{\tr}{Tr}

\newcommand{\numb}{\addtocounter{equation}{1}\tag{\theequation}}

\begin{document}
\title{Exact analytical treatment of multiqubit noisy dynamics in exchange-coupled semiconductor spin qubits}
\author{Donovan Buterakos}
\author{Sankar Das Sarma}
\affiliation{Condensed Matter Theory Center and Joint Quantum Institute, Department of Physics, University of Maryland, College Park, Maryland 20742-4111 USA}
\date{\today}

\begin{abstract}
	Charge noise remains the primary obstacle to the development of quantum information technologies with semiconductor spin qubits. We use an exact analytical calculation to determine the effects of quasistatic charge noise on a ring of three equally spaced exchange-coupled quantum dots. We calculate the disorder-averaged return probability from a specific initial state, and use it to determine the coherence time $T_2^*$ and show that it depends only on the disorder strength and not the mean interaction strength. We also use a perturbative approach to investigate other arrangements of three or four qubits, finding that the return probability contains multiple oscillation frequencies. These oscillations decay in a Gaussian manner, determined by differences in energy levels of the Hamiltonian. We give quantitative values for gate times resulting in several target fidelities. We find that the decoherence time decreases with increasing number of qubits. Our work provides useful analytical insight into the charge noise dynamics of coupled spin qubits.
\end{abstract}

\maketitle

\section{Introduction}

Electron spin qubits in semiconductor (particularly, Si) quantum dots is considered to be a promising platform for developing quantum information technologies due to its long coherence times and fast gate speed, as well as the potential scalability and ability to use resources from the semiconductor industry which are currently available. Semiconductor spin qubits remain among the most-studied quantum computing platforms in the world with laboratories in USA, Australia, Europe, Japan, Canada, and China actively pursuing coupled semiconductor quantum dot systems for eventual quantum computing applications. There has been much experimental progress including the implementation of two qubit gates between singlet-triplet qubits \cite{NicholNPJQI2017}, the fabrication of linear arrays of nine quantum dots \cite{ZajacPRApp2016}, the shuttling of spins across linear arrays \cite{SigillitoNPJQI2019}, and the implementation of basic quantum algorithms on a programmable two-qubit processor \cite{WatsonNAT2018}. A plaquette of four quantum dots has also been used to perform simulations of the Hubbard model and observe Nagaoka ferromagnetism \cite{DehollainNAT2020, WangPRB2019, ButerakosPRB2019}. A four-qubit quantum processor has been constructed with hole qubits in Geranium quantum dots \cite{vanRiggelenAPL2021,HendrickxARXIV2020}. Very recent unpublished work from Delft reports the fabrication of a 6-spin qubit Si system, where our work should be relevant \cite{VandersypenPC}. These are just a few examples from the many recent exciting developments in the field. In spite of much impressive progress, the subject faces a rather difficult challenge in mitigating errors arising from charge noise, which is invariably present in all electronic materials, devices, and circuits.  In particular, charge noise has prevented semiconductor quantum dot platforms from developing multiqubit system operations with the current limit being 2-4 coupled qubits at most.

Benchmarking of quantum dot devices has obtained single qubit gates with fidelities over 99.9\% \cite{YonedaNAT2018}, but two-qubit gates have fidelities ranging from 90\% to 95\%, with a few specific gates reaching up to 98\% \cite{HuangNAT2019,XuePRX2019}. However, the fidelities of these two-qubit gates must still be increased substantially ($>\!99\%$) in order to meet the minimum threshold values needed to implement quantum error correcting codes. The two-qubit gate fidelities are predominantly limited by charge noise, which can arise from local charge impurities in the solid or in the controlling circuits which create the quantum dot potential wells \cite{HuPRL2006,DialPRL2013}. These impurities can affect the electron wave functions, and in turn affect the strength of the exchange interaction used to perform the two-qubit gates, since the exchange interaction depends very precisely on the wavefunction overlap \cite{HuPRL2006}. Methods have been proposed to reduce the effect of charge noise, such as using quadrupolar exchange-only qubits, which uses four electrons in three quantum dots to access a ``sweet spot'' where qubit operation is robust to first order charge noise \cite{RussPRL2018}. Additionally, there are proposals for dynamical decoupling schemes, which perform rotations using complex pulse shapes which cancel errors to some degree \cite{WangNCOM2012,ZengNJP2018}; however, dynamical decoupling requires a precise understanding of the source and form of the noise to be canceled. None of the proposed techniques for mitigation has been generically successful in eliminating charge noise, and understanding and eliminating charge noise remains the main obstacle in the development of semiconductor spin qubits.

The effects of noise have been studied theoretically in coupled double quantum dot systems \cite{ThrockmortonPRB2017,DasSarmaPRB2016,WuPRB2017,WuPRA2016,ThrockmortonPRB2020}. Beginning with a Hamiltonian for two exchange coupled dots in the presence of an external magnetic field, the probability $P(t)$ was calculated that the system initialized in a given state would be measured in the same state after evolution for some time $t$. Quasistatic noise was modeled by choosing the magnetic field and exchange interaction strengths from random Gaussian distributions, and the return probability was analytically averaged over all choices for each parameter. This disorder-averaged return probability showed oscillations within a Gaussian-shaped envelope, and the value of $T_2^*$ for the system was obtained from the decay rate of the envelope \cite{ThrockmortonPRB2017}. Numerical simulations have also been performed for larger systems in the same manner, including a system of two capacitively coupled singlet triplet qubits \cite{DasSarmaPRB2016,WuPRB2017} and ion trap spin chains (which have a similar Hamiltonian and dynamics)\cite{WuPRA2016}. This method has also been used to determine the fidelity of spin transport across spin chains via SWAP gates \cite{ThrockmortonPRB2020}. These works are all, however, purely numerical, and although useful in their own right, these numerical calculations fail to provide analytical insight into the noise dynamics.

In this current work, we {\it analytically} calculate the effects of charge noise on various systems of three and four quantum dots. For a ring of three quantum dots, we perform an exact analytical calculation and obtain expressions for the disorder averaged return probability, as well as the spin expectation values and entanglement entropy for each qubit. We find that the shape of the oscillation envelope is completely independent of the mean exchange interaction strength and depends only on the disorder strength, and we obtain the exact coherence time $T_2^*$ as a function of disorder strength. For other systems, we develop a pertubative approach to calculating the noise-induced decoherence effects. We show that multiple frequencies appear in the resulting expressions, and that these frequencies depend on the differences between energy levels of the noiseless Hamiltonian. These frequencies each decay in a Gaussian-like manner with their own decay rate, and the decay rates are given by the first order noise-induced corrections to the energy levels which produce the frequencies. For each system, we calculate $T_2^*$ and find the times $t_c$ at which the system drops below various fidelity benchmarks. Our analytical theory provides detailed insight into the noise dynamics which are hidden (and therefore difficult to discern) in the existing numerical calculations. It is in fact quite an unanticipated finding that the multiqubit dynamics under charge noise can be obtained analytically.

This paper is organized as follows: in Sec. II, we examine a ring of three equally spaced quantum dots, and perform an exact analytical calculation of several quantities in the presence of noise, including the disorder averaged return probability from a specific initial state. We give an analysis of the results, discussing the short and long time behavior of the system and its dependence on the initial Hamiltonian. In Sec. III, we address a system of three quantum dots in a linear geometry by using a perturbative approach, and we compare our results to a direct numerical evaluation of the same quantities. In Sec IV, we use the same perturbative approach to examine a ring of four quantum dots. Finally, we conclude with a discussion in Sec. V.

\section{Three Qubit Ring}

In this section we calculate the effect of noise on a quantum dot plaquette with three qubits arranged in an equilateral triangle. Specifically, we consider the case where the mean interaction strength between each pair of dots is identical. This case has a high degree of symmetry, which allows the dynamics in the presence of noise to be calculated exactly analytically. We begin by defining the model and Hamiltonian; we then show the calculation of the disorder averaged return probability from an initial state; finally, we discuss the results.

\subsection{Model and Hamiltonian}

We consider a single half-filled band in a plaquette consisting of a ring of three quantum dot, and thus three electrons are present. We neglect the effect of higher unoccupied energy orbitals as the energy difference between bands tends to be much larger than the interaction strength between dots, which is the relevant energy scale for the dynamics of the system. Similarly, we also assume a large onsite interaction energy, and thus we ignore states where a single dot contains more than one electron (these are all experimentally valid and theoretically used approximations for semiconductor quantum dot qubits). Then each dot will contain exactly one electron, and the exchange interaction between dots will give rise to the following Heisenberg Hamiltonian:

\begin{equation}
H=J_{12}\vec{S}_1\cdot\vec{S}_2+J_{23}\vec{S}_2\cdot\vec{S}_3+J_{13}\vec{S}_1\cdot\vec{S}_3
\label{eqn:h3tri}
\end{equation}

Let the values of the exchange interaction between two dots $J_{ij}$ have a Gaussian distribution because of the noise with mean $J_0$ and standard deviation $\sigma_J$. Here, $\sigma_J$ is a measure of the charge noise induced disorder, leading to random Gaussian fluctuations in the exchange coupling. In order for this distribution to be sensible, $J_0$ must be several times larger than $\sigma_J$, so that the proportion of the distribution with a negative value of $J$ is negligible, and we will assume this is the case. It is convenient to define and work with the deviations $\Delta_{ij}=J_{ij}-J_0$. These deviations will then have the following distribution

\begin{equation}
f(\Delta)=\frac{1}{\sqrt{2\pi}\sigma_J}e^{-\frac{\Delta^2}{2\sigma_J^2}}
\end{equation}

For any function or operator $A$ which depends on $J_{ij}$, we define its disorder averaged expectation $[A]$ as follows:

\begin{equation}
[A]=\int A\;f(\Delta_{12})f(\Delta_{23})f(\Delta_{13})\;d\Delta_{12}d\Delta_{23}d\Delta_{13}
\label{eqn:aave}
\end{equation}

Note that the Hamiltonian in Eq. (\ref{eqn:h3tri}) commutes with the total spin operator $S^2$, as well as its z-projection, and thus both quantities are conserved. There is only one state with $S_z=3/2$, and thus its dynamics is trivial. We will focus on the dynamics of the $S_z=1/2$ subspace.

Our model relies on several assumptions. We assume a linear noise term, and thus our results may not apply to select systems with quadratic noise terms, such as in Ref. \onlinecite{RussPRL2018}, although the majority of experiments do exhibit linear noise. It should be possible to generalize our work to different noise terms, but this is beyond the scope of the current work where we establish the general theory and principles using the usual linear noise model. Additionally, we use the quasistatic noise approximation, which is well-justified in quantum dot systems since gate times are very fast compared to noise frequencies. In fact, the whole point of semiconductor spin qubits is very high gate speed and the prospect for future scalability. We also exclude the Zeeman interaction term in our model, as our focus for this paper is charge noise. In fact, it is well-known that field noise arising from nuclear spin coupling is strongly suppressed in Si and can even be eliminated by isotopic purification, making Si spin the ideal semiconductor material for qubit applications \cite{WitzelPRL2010}, leaving charge noise as the only operational decohering mechanism. The effect of the Zeeman interaction on qubit dynamics has been studied extensively in other works \cite{ThrockmortonPRB2017,DasSarmaPRB2016,WuPRB2017}, and are small when the exchange interaction terms $J_{ij}$ are large, which is when there is the greatest amount of charge noise. In Si spin qubits, field noise is far too weak for it to affect qubit operations at the operational gate frequencies. Additionally, some qubit proposals do not have a Zeeman term present at all, with both one- and two-qubit operations performed via exchange coupling \cite{DiVincenzoNAT2000}. Various other system-dependent deviations from our model may occur, such as changes to the exchange interaction due to valley splitting in Si. However, as long as such effects are small, the resulting dynamics will still be qualitatively the same. Quantitative numerics require a precise knowledge of the microscopic Hamiltonian, which is system-dependent and generally unknown, and thus we study the qualitative behavior of the system. Our current theory should be thought of as the minimal theory which must be the starting point for understanding noisy qubit dynamics in semiconductor spin systems.

\begin{figure*}[!tb]
	
	\includegraphics[width=.9\columnwidth]{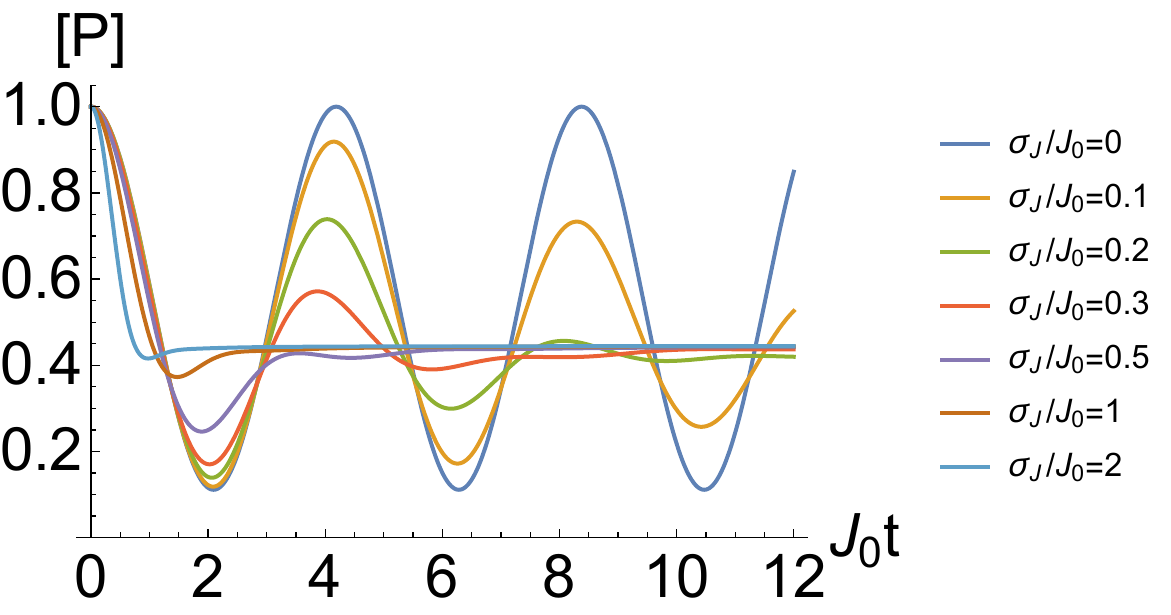}
	\includegraphics[width=.9\columnwidth]{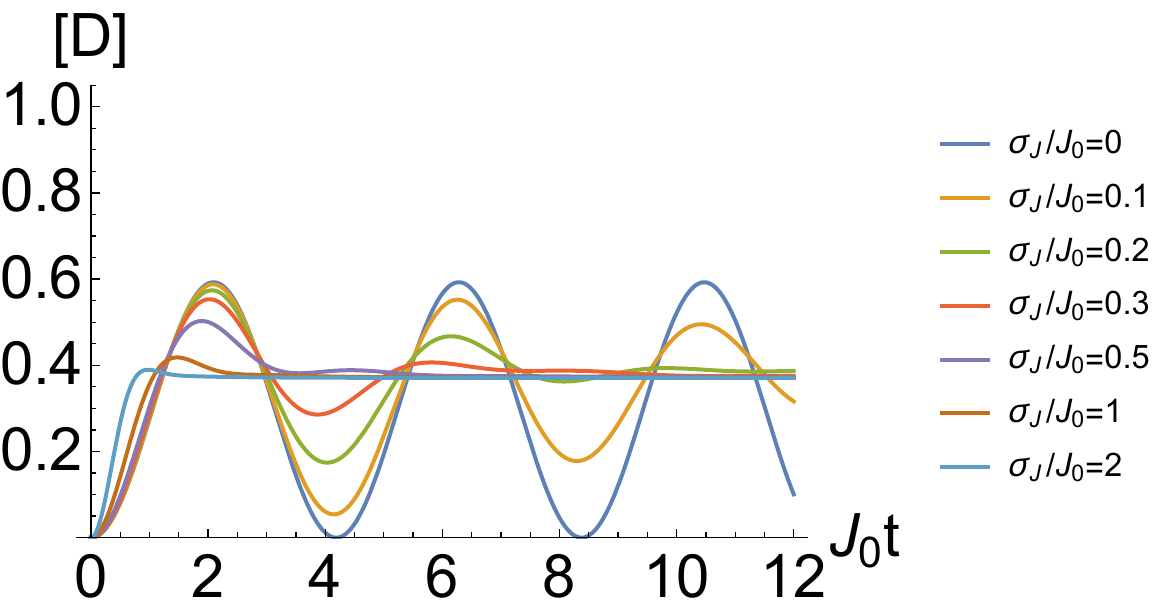}
	\includegraphics[width=.9\columnwidth]{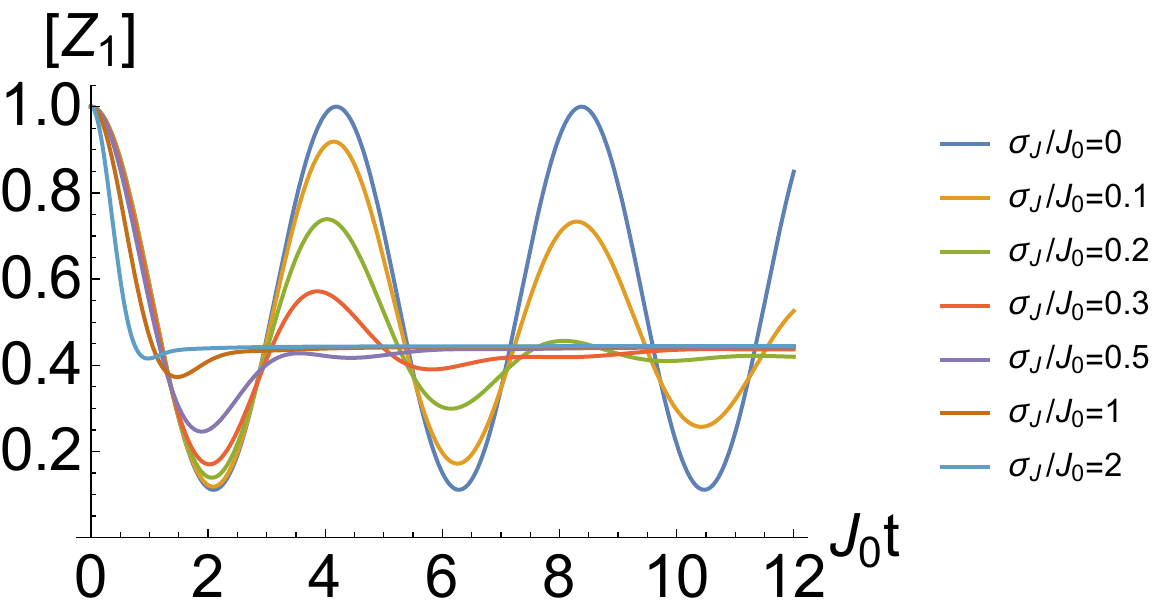}
	\includegraphics[width=.9\columnwidth]{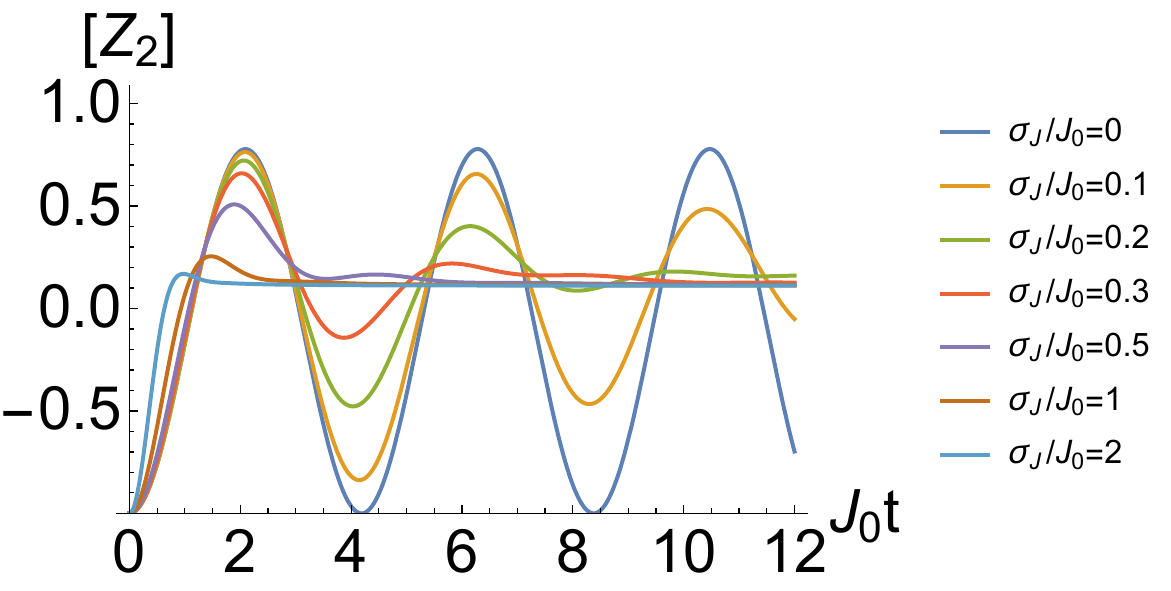}
	\includegraphics[width=.9\columnwidth]{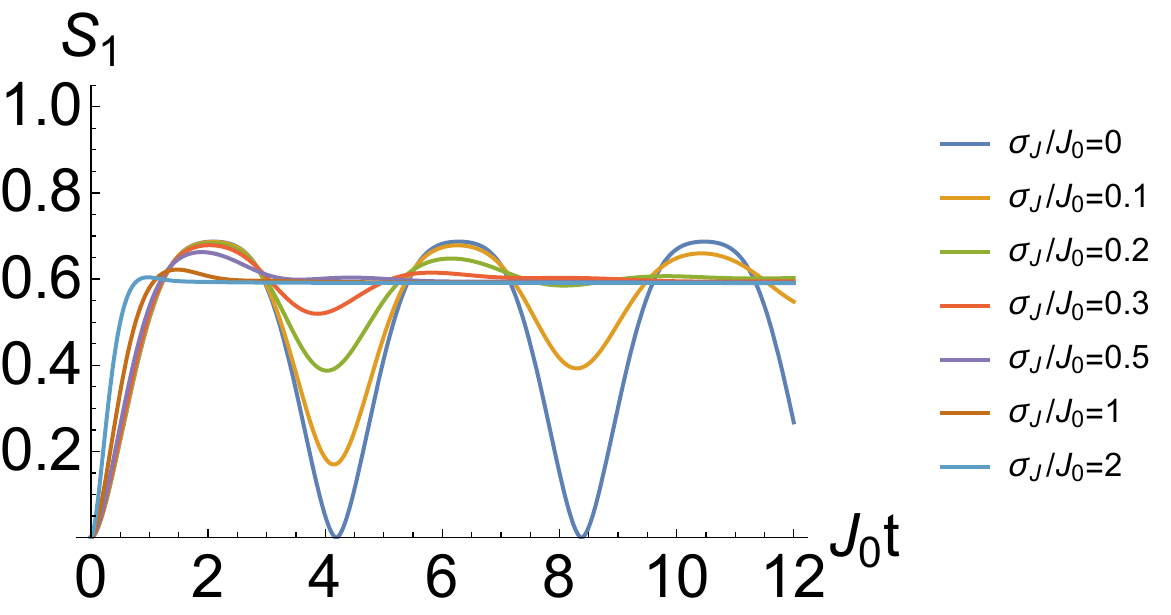}
	\includegraphics[width=.9\columnwidth]{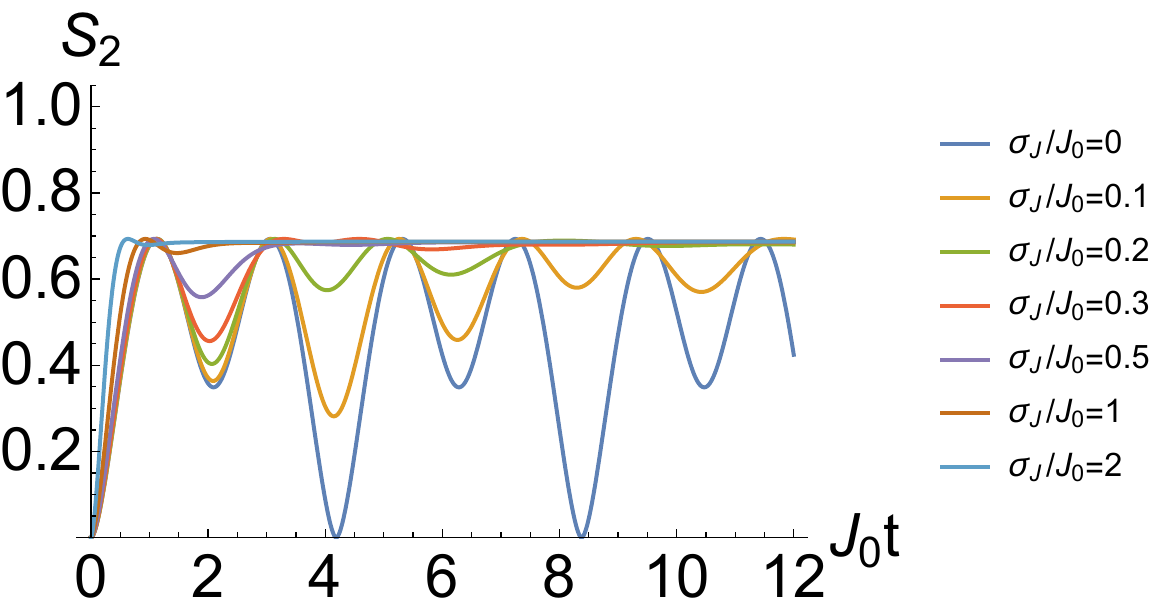}
	\caption{The disorder averaged return probability $[P(t)]$, normalized Hamming distance $[D(t)]$, expectation values $[Z_1(t)]$ and $[Z_2(t)]$, and entanglement entropy of qubits 1 \& 2 $[S_1(t)]$ and $[S_1(t)]$. Each is plotted for several values of $\sigma_J$, with $J_0$ held constant.}
	\label{fig:plot3tri}
\end{figure*}

\subsection{Calculation of Return Probability}

We consider the system prepared in an initial state $\ket{\ua\da\ua}$, and define the return probability $P(t)$ to be the probability of measuring the system to be in the same state after it has evolved for some time $t$. Note that this initial state is the only nontrivial state with individual spins initialized to $\ua$ or $\da$, since other combinations are equivalent by symmetry. There are three states with total $S_z=1/2$: $\{\ket{\ua\ua\da}$, $\ket{\ua\da\ua}$, and $\ket{\da\ua\ua}\}$. These are not spin eigenstates, so we instead use the following basis:

\begin{equation}
\ket{\phi_j}=\frac{e^{2j\pi i/3}\ket{\ua\ua\da}+\ket{\ua\da\ua}+e^{-2j\pi i/3}\ket{\da\ua\ua}}{\sqrt{3}}
\end{equation}

for $j=-1,0,1$, where $\ket{\phi_0}$ has spin $3/2$, and $\ket{\phi_{\pm1}}$ have spin $1/2$. In this basis, $H$ is given by:

\begin{align*}
&H=\frac{3J_0}{4}\begin{pmatrix}-1&0&0\\0&1&0\\0&0&-1\end{pmatrix}+\frac{1}{4}\begin{pmatrix}
-\Sigma_\Delta&0&2\xi_\Delta^*\\0&\Sigma_\Delta&0\\2\xi_\Delta&0&-\Sigma_\Delta
\end{pmatrix}
\numb
\end{align*}

where $\Sigma_\Delta=\Delta_{12}+\Delta_{13}+\Delta_{23}$, and $\xi_\Delta=e^{2\pi i/3}\Delta_{12}+\Delta_{13}+e^{-2\pi i/3}\Delta_{23}$. Diagonalizing gives the following energies and eigenvectors:

\begin{align*}
E_0&=\frac{3J_0}{4}+\frac{\Sigma_\Delta}{4}\\
E_\pm&=-\frac{3J_0}{4}-\frac{\Sigma_\Delta}{4}\pm\frac{|\xi_\Delta|}{2}\\
\ket{\psi_0}&=\ket{\phi_0}\\
\ket{\psi_\pm}&=\frac{\pm\chi_\Delta^*}{{\sqrt{2}}}\ket{\phi_{-1}}+\frac{1}{\sqrt{2}}\ket{\phi_{+1}}
\numb
\end{align*}

where $\chi_\Delta=\xi_\Delta/|\xi_\Delta|$. The initial state $\ket{\ua\da\ua}$ corresponds to the vector $(1\;1\;1)/\sqrt{3}$ in the $\phi_j$ basis, and decomposing into the basis of eigenstates, this becomes:

\begin{equation}
\ket{\ua\da\ua}=\frac{1-\chi_\Delta}{\sqrt{6}}\ket{\psi_-}+\frac{1}{\sqrt{3}}\ket{\psi_0}+\frac{1+\chi_\Delta}{\sqrt{6}}\ket{\psi_+}
\end{equation}

We then calculate the return probability $P(t)$ for a particular disorder realization, yielding:

\begin{align*}
&P(t)=\frac{1}{9}\Big(2\cos \frac{|\xi_\Delta|}{2}t+\cos\frac{3J_0+\Sigma_\Delta}{2}t\Big)^2\\&\qquad+\frac{1}{9}\Big(2\re\chi_\Delta\sin \frac{|\xi_\Delta|}{2}t+\sin\frac{3J_0+\Sigma_\Delta}{2}t\Big)^2
\numb
\label{eqn:p}
\end{align*}

From this expression, the disorder average $[P(t)]$ can be calculated using eq. (\ref{eqn:aave}). In order to compute the integral, it is helpful to change the variables of integration from $\Delta_{ij}$ to $\Sigma_\Delta$ and $\xi_\Delta$, the latter of which takes values over the whole complex plane. We then write $\xi_\Delta$ in terms of its magnitude and complex argument $\varphi_\Delta=\arg\xi_\Delta$, which produces the following integral:

\begin{align*}
&[P(t)]=\int P(t)\;\frac{e^{-(\Sigma_\Delta^2+2|\xi_\Delta|^2)/6\sigma_J^2}}{(\sqrt{2\pi}\sigma_J)^3}\;\frac{2|\xi_\Delta|}{3\sqrt{3}}\;d\Sigma_\Delta d|\xi_\Delta|d\varphi_\Delta
\numb
\label{eqn:pint}
\end{align*}

This integral can be evaluated and expressed in terms of the Dawson function $F(x)$, as follows:

\begin{align*}
[P(t)]&=\frac{5}{9}+\frac{4}{9}e^{-3\sigma_J^2t^2/8}\cos\frac{3J_0t}{2}-\frac{1}{3\sqrt{3}}\sigma_JtF\Big(\frac{\sqrt{3}\sigma_Jt}{2}\Big)\\&-\frac{2}{3\sqrt{3}}e^{-3\sigma_J^2t^2/8}\sigma_Jt\cos\frac{3J_0t}{2}F\Big(\frac{\sqrt{3}\sigma_Jt}{4}\Big)
\numb
\label{eqn:pave}
\end{align*}

Using a similar process, we calculate the disorder averaged expectation value of the $Z_2$ operator (the Pauli Z operator on the qubit 2), yielding:

\begin{align*}
&[Z_2]=\frac{1}{9}\bigg(-1+2\sqrt{3}\sigma_JtF\big(\frac{\sqrt{3}\sigma_Jt}{2}\big)\\&+4e^{-3\sigma_J^2t^2/8}\cos\frac{3J_0t}{2}\Big(-2+\sqrt{3}\sigma_JtF\big(\frac{\sqrt{3}\sigma_Jt}{4}\big)\Big)\bigg)
\numb
\end{align*}

The expectation values of the other two spins $[Z_1]=[Z_3]$ can be obtained directly from this result, since the three must sum to 1. We plot these expectation values as a function of time with $J_0$ constant in Fig. \ref{fig:plot3tri}. These expectation values can be used to calculate the expectation of the Hamming distance from the initial to final states. The Hamming distance is a quantity used for error-correction which measures the number of qubits which must be flipped to transition between the two states, and thus is a measure of the error introduced into the system \cite{WuPRA2016,HammingBSTJ1950}. In in Fig. \ref{fig:plot3tri} we plot normalized Hamming distance defined by:

\begin{equation}
D(t)=\frac{1}{2}+\frac{1}{6}\Big(\langle Z_1\rangle-\langle Z_2\rangle+\langle Z_3\rangle\Big)
\end{equation}

Additionally, in Fig. \ref{fig:plot3tri} we plot the entanglement entropy between qubit $j$ and the rest of the system, defined to as:

\begin{equation}
S_j=-\tr\rho_j\log\rho_j
\end{equation}

where $\rho(t)$ is the disorder-averaged density matrix of the system evolved a time $t$ from the initial state $\ket{\uparrow\downarrow\uparrow}$, and $\rho_j$ is $\rho$ traced over all qubits except qubit $j$.

\begin{figure*}[!tb]
	
	\includegraphics[width=.9\columnwidth]{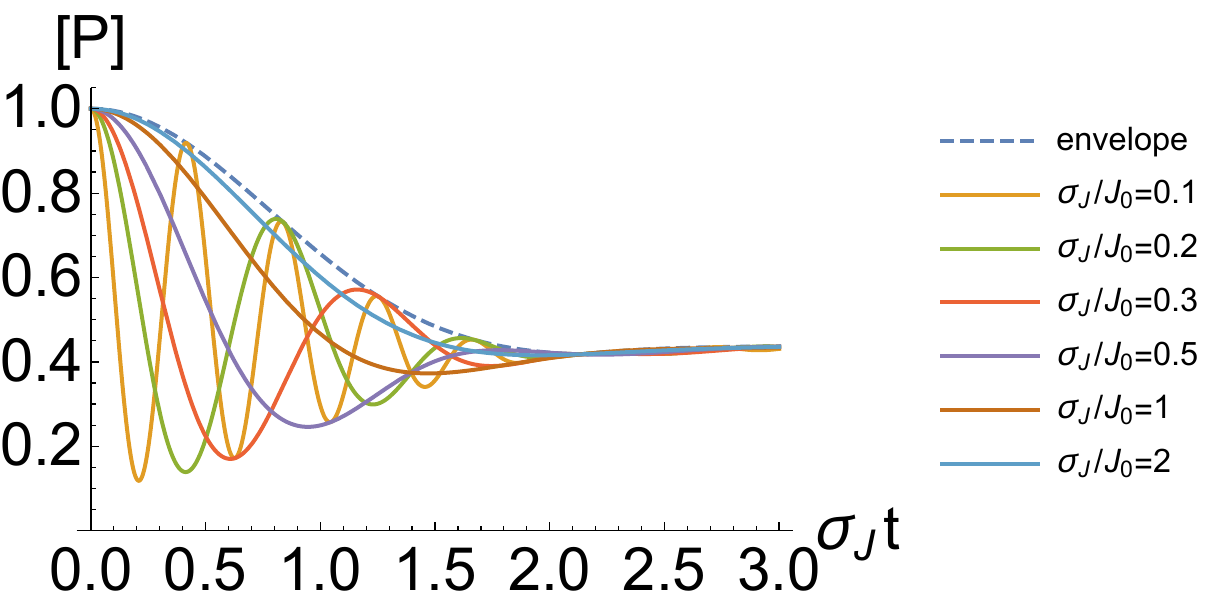}
	\includegraphics[width=.9\columnwidth]{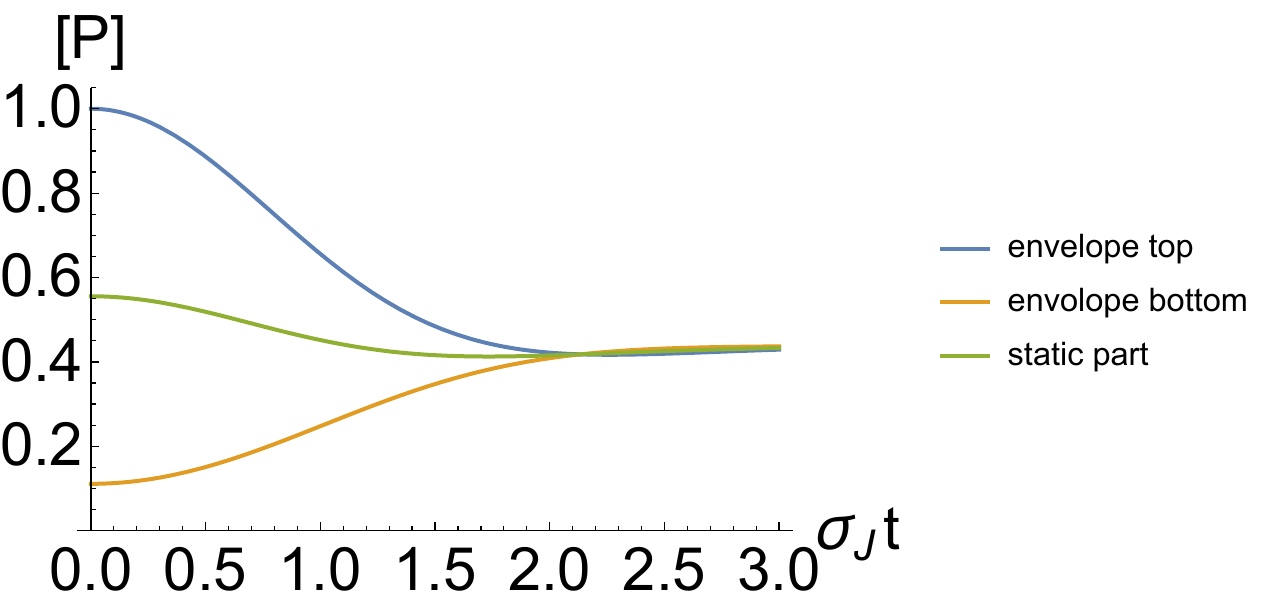}
	\includegraphics[width=.9\columnwidth]{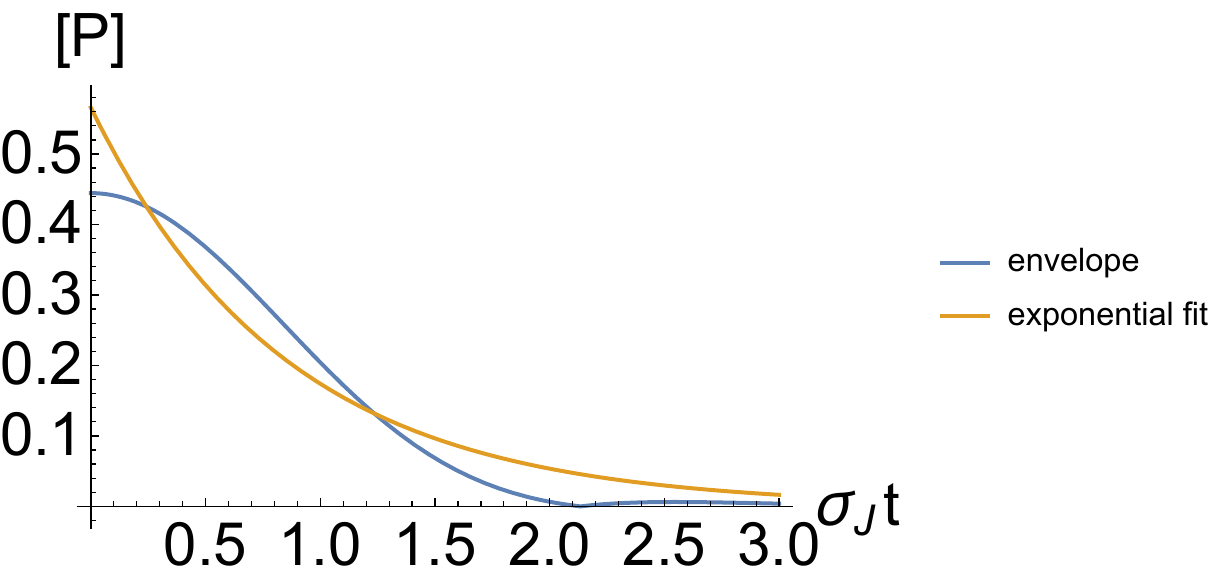}
	\includegraphics[width=.9\columnwidth]{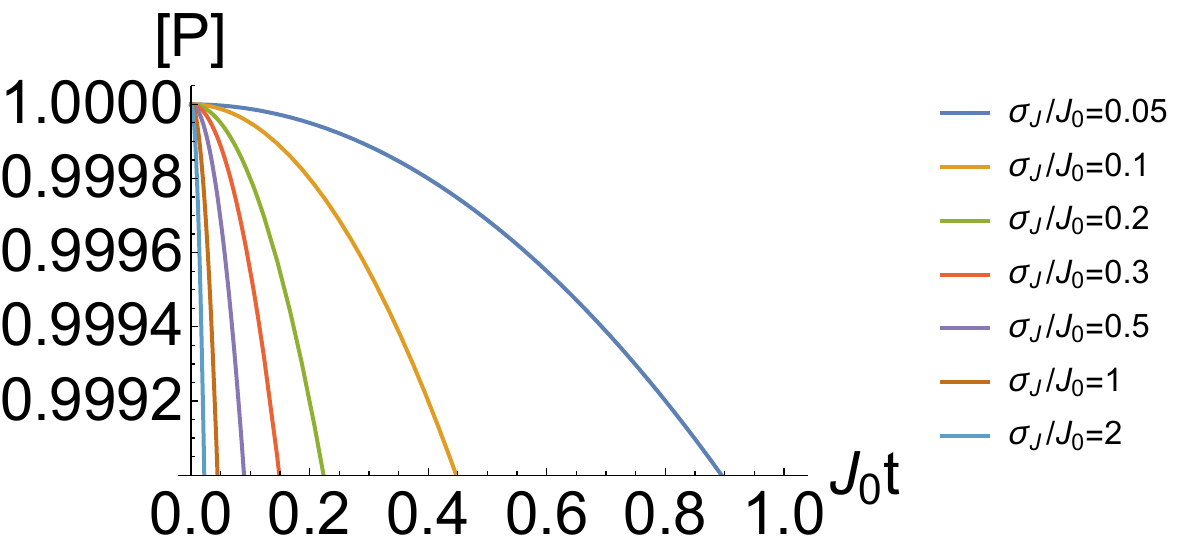}
	\caption{{\bf Top Left:} The return probability plotted against time with $\sigma_J$ held constant. {\bf Top Right:} The oscillation envelope plotted alongside the static part $[P(t)]_{\text{st}}$. Note the node at $t=2.13\sigma_J^{-1}$. {\bf Bottom Left:} The oscillatory part of the envelope and its closest exponential fit. {\bf Bottom Right:} The return probability plotted for small $t$, with $J_0$ held constant.}
	\label{fig:env3tri}
\end{figure*}

\subsection{Analysis}

We examine the behavior of the return probability $[P(t)]$ given by Eq. (\ref{eqn:pave}) by separating $[P(t)]$ into its static and oscillatory parts. We define static part as the midpoint of the oscillation envelope as a function of time, and the oscillatory part of $[P(t)]$ is the deviation of $[P(t)]$ from the envelope's midpoint, which is thus determined by the envelope's width. Expanding in $(\sigma_J t)^{-1}$, we find the long-time asymptotic behavior of the static and oscillatory parts as follows, which we plot in Fig. \ref{fig:env3tri}:

\begin{align}
&[P(t)]_{\text{st}}=\frac{4}{9}-\frac{2}{27\sigma_J^2t^2}+O\bigg(\frac{1}{\sigma_J^4t^4}\bigg)\nonumber\\
&[P(t)]_{\text{osc}}=e^{-3\sigma_J^2t^2/8}\cos\frac{3J_0t}{2}\bigg[-\frac{32}{27\sigma_J^2t^2}+O\Big(\frac{1}{\sigma_J^4t^4}\Big)\bigg]
\end{align}

For $t\gg\sigma_J^{-1}$, the static part varies as $t^{-2}$, but the oscillatory part falls off as $e^{-t^2}$, so oscillations are only detectable on short time scales. Specifically, we note that the Gaussian prefactor for the oscillatory part of the envelope implies that it decays much faster than the standard exponential approximation often associated with noise. It is interesting to note that the coefficient of the oscillatory part of the envelope vanishes at the point $\sigma_Jt=\frac{4}{\sqrt{3}}F^{-1}(\frac{2}{\sqrt{3}})\approx2.13$, and the coefficient changes sign as $t$ crosses this point. This means that there will always be a node at this point due to the shape of the envelope, which can be seen in Fig. \ref{fig:env3tri}. We also give the short-time behavior by expanding $[P(t)]$ with $t\ll\sigma_J^{-1}$, yielding the following:

\begin{equation}
[P(t)]=\bigg(\frac{5}{9}-\frac{\sigma_J^2t^2}{6}\bigg)+\bigg(\frac{4}{9}-\frac{\sigma_J^2t^2}{3}\bigg)\cos\frac{3J_0 t}{2}+O(\sigma_J^4t^4)
\label{eqn:paveexpand0}
\end{equation}

Because coherence time $T_2^*$ is formally defined in terms of the decay constant of an exponential curve, it is often implied that noise induces an exponential decay of coherence. In Ref. \cite{ThrockmortonPRB2017} it was shown that quasistatic charge noise in a two-level system produces a Gaussian decay rather than an exponential decay. We extend that result to our system of three qubits, where the oscillation envelope for the return probability is also Gaussian-like in nature, with its exact functional form given in Eq. (\ref{eqn:pave}). The exponential decay approximation simply is invalid here. In Fig. \ref{fig:env3tri}, we show the half width of the oscillation envelope alongside its closest exponential fit using a least squares fit. Note that the exponential curve is a very poor approximation of the oscillation envelope. Thus, we define $T_2^*$ to be the point where the envelope width reaches $1/e$ of its original value, and we stress that this only gives the relative time scale on which the system decoheres and does not imply exponential behavior. This $T_2^*$ is simply an operational coherence time. Having a well-defined effective `coherence time' $T_2^*$ is not dependent on the actual details of the decoherence process (i.e. exponential versus Gaussian) as long as it is clearly defined as we do here.  Experiments are operationally easier to characterize and describe by a single phenomenological parameter such as $T_2^*$, and this is typically done independent of the details of the decoherence mechanism.  Our definition of $T_2^*$ does not make any assumption about the decay of coherence (exponential, Gaussian, or any other functional form) in our system, which we calculate exactly. Note that the oscillation envelope is asymmetric due to the $t F(t)$ term in Eq. (\ref{eqn:pave}), and thus defining $T_2^*$ based on the envelope width versus the total envelope height will give slightly different results, and either is perfectly acceptable as long as it is made clear how $T_2^*$ is being defined. We choose to define $T_2^*$ based on the envelope width, and by this definition, $T_2^*$ depends only on the oscillatory part of $[P(t)]$.

Using this definition, $T_2^*$ is given by:

\begin{equation}
T_2^*=1.127\sigma_J^{-1}
\label{eqn:t2}
\end{equation}

It is interesting that $T_2^*$ is defined only by $\sigma_J$ with $J_0$ playing no role. Note from Eq. (\ref{eqn:pave}) that the only effect that $J_0$ has on the return probability $[P(t)]$ is to set the frequency of oscillations. The shape of the oscillation envelope itself is completely independent of $J_0$. This is evident in Fig. \ref{fig:env3tri}, where curves with different ratios of $\sigma_J/J_0$ fill exactly the same envelope. This phenomenon is due to the symmetry of the system. Specifically, for a pair of qubits, an equilateral triangle, or any number of qubits with a complete graph of equal exchange interactions between every pair of qubits, the unperturbed Hamiltonian $H_0$ will be proportional to the total spin operator $S^2$. In order for the shape of the envelope to be affected by $J_0$, there must be some noise term that mixes states with energy difference on the order of $J_0$. However, $S^2$ is conserved regardless of the disorder realization, and thus only states with the same spin (and therefore the same energy under $H_0$) can mix. In order for $J_0$ to affect the shape of the oscillation envelope, it is necessary that the unperturbed Hamiltonian have eigenstates with the same spin and differing energies.

\section{3-qubit Linear Array}

To demonstrate how qubit geometry affects decoherence, we now consider a linear array of three dots in the open geometry, given by the following Hamiltonian:

\begin{equation}
H=J_{12}\vec{S}_1\cdot\vec{S}_2+J_{23}\vec{S}_2\cdot\vec{S}_3
\label{eqn:h3lin}
\end{equation}

Because the system is lacking the same symmetry as before, an exact analytical expression for $[P(t)]$ is not obtainable. We approach the problem in two ways: first, analytically by using perturbation theory for $\sigma_J\ll J_0$, and second by showing the results of direct numerics.

\subsection{Perturbative Approach}

We begin with the unperturbed Hamiltonian $H_0$ given by Eq. (\ref{eqn:h3lin}) with $J_{12}=J_{23}=J_0$. We then add perturbations $\Delta_{ij}=J_{ij}-J_0$, and find the energies of the eigenstates to order $\Delta$. We also find the overlap $a_i$ of each eigenstate with the initial state $\ket{\ua\da\ua}$. These are given by:

\begin{align*}
&E_1=\frac{J_0}{2}+\frac{\Delta_{12}+\Delta_{23}}{4}+O\Big(\frac{\Delta^2}{J_0}\Big)\\
&E_2=O\Big(\frac{\Delta^2}{J_0}\Big)\\
&E_3=-J_0-\frac{\Delta_{12}+\Delta_{23}}{2}+O\Big(\frac{\Delta^2}{J_0}\Big)\\
&|a_1|^2=\frac{1}{3}+O\Big(\frac{\Delta^3}{J_0^3}\Big)\\
&|a_2|^2=\frac{(\Delta_{12}-\Delta_{23})^2}{8J_0^2}+O\Big(\frac{\Delta^3}{J_0^3}\Big)\\
&|a_3|^2=\frac{2}{3}-\frac{(\Delta_{12}-\Delta_{23})^2}{8J_0^2}+O\Big(\frac{\Delta^3}{J_0^3}\Big)
\numb
\end{align*}

The return probability for a particular disorder realization is given by:

\begin{equation}
P(t)=\bigg|\sum_n |a_n|^2e^{-iE_nt}\bigg|^2=\sum_{n,m}|a_na_m|^2\cos(E_n-E_m)t
\label{eqn:pgen}
\end{equation}

Then calculating the disorder average as in Eq. (\ref{eqn:aave}), we obtain $[P(t)]$ as follows:

\begin{align*}
&[P(t)]=\frac{5}{9}-\frac{\sigma_J^2}{3J_0^2}+e^{-\sigma_J^2t^2/16}\frac{\sigma_J^2}{6J_0^2}\cos\frac{J_0t}{2}\\&+e^{-\sigma_J^2t^2/4}\frac{\sigma_J^2}{3J_0^2}\cos J_0t\\&+e^{-9\sigma_J^2t^2/16}\Big(\frac{4}{9}-\frac{\sigma_J^2}{6J_0^2}\Big)\cos\frac{3J_0t}{2}+O\Big(\frac{\sigma_J^3}{J_0^3},\frac{\sigma_J^2t}{J_0}\Big)
\numb
\label{eqn:pave3lapr}
\end{align*}

In the limit where $J_0/\sigma_J\rightarrow\infty$, this reduces to $\frac{5}{9}+\frac{4}{9}e^{-9\sigma_J^2t^2/16}\cos\frac{3J_0t}{2}$, and thus the time $T_2^*$ for the envelope width to drop to $1/e$ times its initial width is given by:

\begin{equation}
T_2^*=\frac{4}{3}\sigma_J^{-1},
\end{equation}

which is within 20\% of the corresponding $T_2^*$ in Eq. (\ref{eqn:t2}) for the triangular ring arrangement of the qubits. Note that there are three frequencies present in the expression for $[P(t)]$. In general, these correspond to the energy differences between pairs of eigenstates of the unperturbed Hamiltonian $H_0$. The Gaussian decay factor for each frequency is determined by the first-order correction to the energies. We show this by letting the cosine term of Eq. (\ref{eqn:pgen}) take the form:

\begin{align*}
&\cos(E_n-E_m)t=\cos(\delta J+c_{12}\Delta_{12}+c_{23}\Delta_{23})t\\
&=\cos\delta J\,t\,\cos(c_{12}\Delta_{12}+c_{23}\Delta_{23})t\\&-\sin\delta J\,t\,\sin(c_{12}\Delta_{12}+c_{23}\Delta_{23})t
\numb
\end{align*}

Then for each variable $\Delta$ integrated over to compute the disorder average, the $\cos c\Delta t$ term produces a Gaussian factor of $e^{-c^2t^2/2\sigma_J^2}$. There is physical meaning behind the connection of these factors to the energies of the system. Because the oscillation frequencies are determined by the unperturbed energies, a disordered system will oscillate with the same frequencies as a clean system,  independent of the disorder strength present. However, the rate at which these oscillations decay is determined by the sensitivity of these energies to disorder. Energy gaps which are less affected by disorder will produce longer-lasting oscillations. This is the essence of noise-induced decoherence in general.

\begin{figure}[!htb]
	\includegraphics[width=\columnwidth]{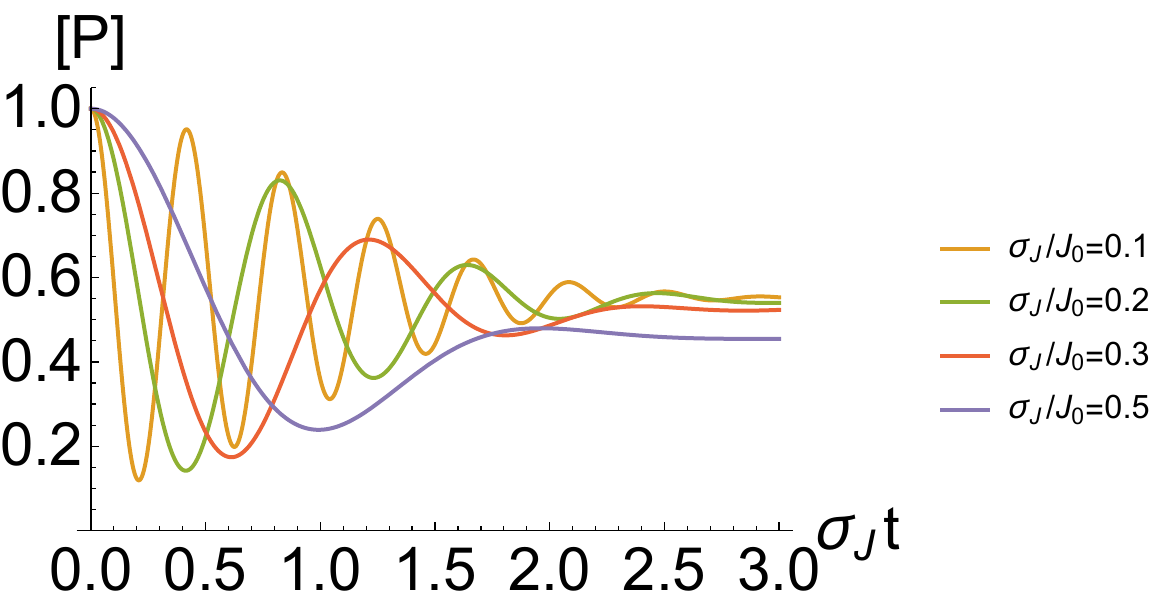}
	\includegraphics[width=\columnwidth]{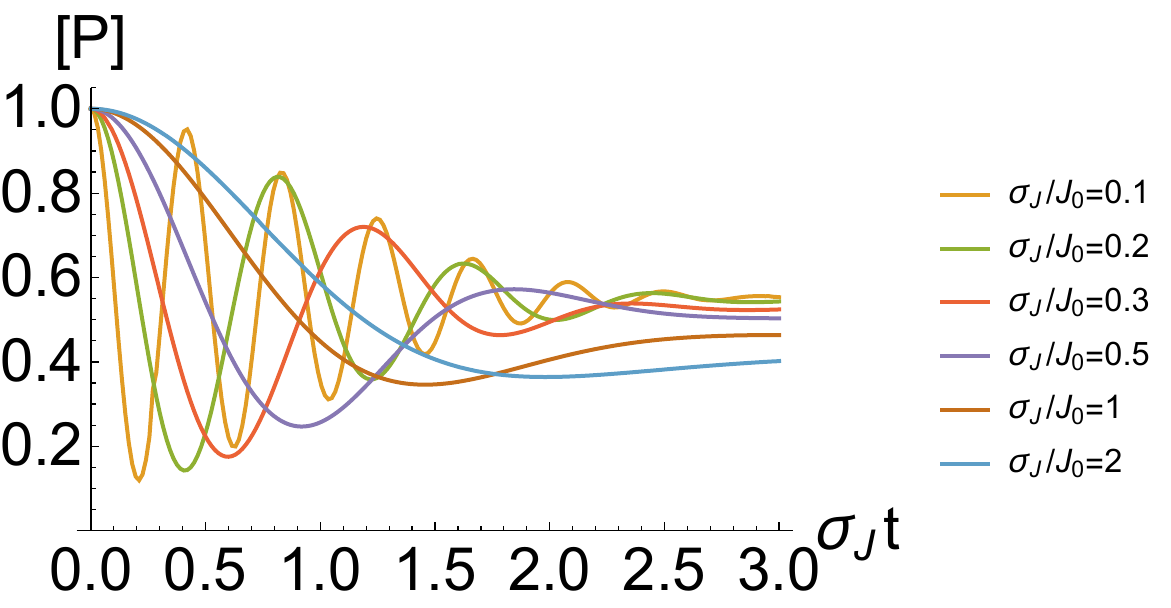}
	\caption{{\bf Top:} The expectation of return probability $[P(t)]$ for different values of $\sigma_J$, calculated from the perturbative result given by eq. (\ref{eqn:pave3lapr}). {\bf Bottom:} The same quantities calculated exactly with numerics from Eq. (\ref{eqn:numint}).}
	\label{fig:pave3l}
\end{figure}

\subsection{Numerical Results}

An alternative approach is to use the same expressions as the case in Sec. II with the equilateral triangle, but enforcing $J_{13}=0$ to be consistent with topology of the linear chain. This will correspond to setting $\Delta_{13}=-J_0$ throughout the calculation. Then we arrive at a corresponding integral to Eq. (\ref{eqn:pint}):

\begin{align*}
&\int P(t)\frac{1}{(\sqrt{2\pi}\sigma_J)^2}e^{-(\Delta_1^2+\Delta_3^2)/2\sigma_J^2}\delta(\Delta_2+J_0)d\Delta_1d\Delta_2d\Delta_3\\
&=\int P(t)\frac{1}{(\sqrt{2\pi}\sigma_J)^2}e^{-(\Sigma_\Delta^2+2|\xi_\Delta|^2-3J_0^2)/6\sigma_J^2}\frac{2|\xi_\Delta|}{3\sqrt{3}}\times\\&\qquad\delta\bigg(\frac{\Sigma_\Delta+2|\xi_\Delta|\cos\varphi_\Delta}{3}+J_0\bigg)d\Sigma_\Delta d|\xi_\Delta|d\varphi_\Delta
\numb
\label{eqn:numint}
\end{align*}

where $P(t)$ is given by Eq. (\ref{eqn:p}). This integral is difficult to evaluate analytically, but can be done numerically. In Fig. \ref{fig:pave3l}, we show the plot of $[P(t)]$ obtained perturbatively from Eq. (\ref{eqn:pave3lapr}) alongside the same numerical results. For $\sigma_J/J_0$ of $0.1$ and $0.2$, the plots are nearly identical; however, significant differences can be seen for larger values such as $0.5$, where the assumption $\sigma_J\ll J_0$ is no longer valid, making the perturbation theory inaccurate. From the numerical results, we note that for different values of $J_0$, the oscillation envelopes tend to be mostly the same in shape. However, there is some small dependence on $J_0$ in the linear chain, as is most evident by the fact that the asymptotes of $[P(t)]$ are different as $J_0$ differs. Additionally, comparing the locations of the peaks in Fig. \ref{fig:pave3l}, we see that the peaks corresponding to $\sigma_J/J_0=0.1$ (yellow) are slightly higher than the corresponding peaks for other values of $\sigma_J/J_0$. This demonstrates that the strict independence of $[P(t)]$ from $J_0$ observed in the case of the equilateral triangle above is a result of the symmetry of the system, as breaking this symmetry causes some deviance of the envelopes with $J_0$. However, the results are still reasonably independent of $J_0$ making the analytical results of the triangular ring model approximately applicable even in the absence of the ring symmetry.

\begin{figure}[tbh]
	\includegraphics[width=\columnwidth]{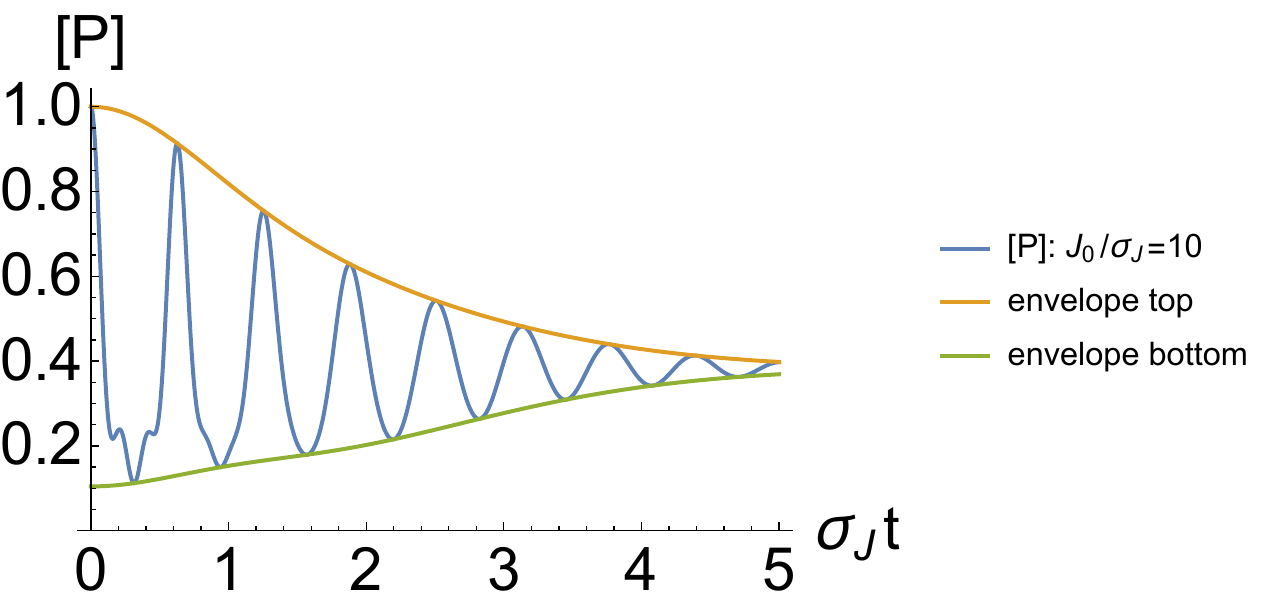}
	\includegraphics[width=\columnwidth]{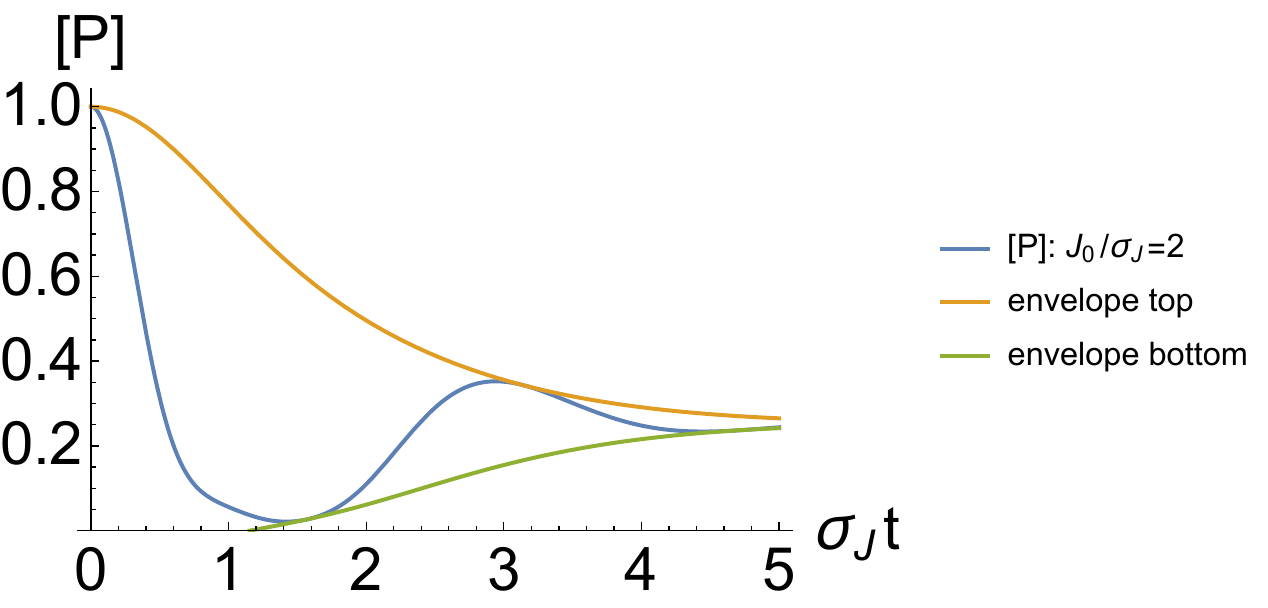}
	\caption{The expectation of return probability $[P(t)]$ given by eq. (\ref{eqn:pave4q}) for $J_0/\sigma_J=10$ (Top) and $J_0/\sigma_J=2$ (Bottom).}
\end{figure}

\section{4-Qubit Square}

\begin{table*}[t]
	\begin{tabular}{ |c|c|c|c|c| } 
		\hline
		fraction & 2-qubit $t_c\sigma_J$ & 3-qubit ring $t_c\sigma_J$ & 3-qubit line $t_c\sigma_J$ &  4-qubit ring $t_c\sigma_J$ \\\hline
		0.9999 & 0.0141 & 0.0115 & 0.0133 & 0.0163 \\\hline 
		0.999 & 0.0447 & 0.0365 & 0.0422 & 0.0517 \\\hline
		0.99 & 0.142 & 0.116 & 0.134 & 0.164 \\\hline
		0.9 & 0.459 & 0.374 & 0.433 & 0.550 \\\hline
		1/e & 1.41 & 1.127 & 1.333 & 2.389 \\\hline
	\end{tabular}
	\caption{Decay rates for the oscillation envelopes for systems of various numbers of qubits. The values in the table give the time $t_c$, in units of $\sigma_J^{-1}$, for the envelope width to decay to a given fraction of its original value. The fraction $1/e$ corresponds to the value of $T_2^*$ for the given system. For the 3-qubit linear array and the 4-qubit ring, values are given for the limit $J_0/\sigma_J\rightarrow\infty$.}
	\label{tab:decay}
\end{table*}

For comparison, we also consider a system of four dots in a square configuration, which gives the following Hamiltonian:

\begin{equation}
H=J_{12}\vec{S}_1\cdot\vec{S}_2+J_{23}\vec{S}_2\cdot\vec{S}_3+J_{34}\vec{S}_3\cdot\vec{S}_4+J_{41}\vec{S}_4\cdot\vec{S}_1
\end{equation}

There is some choice of initial state, but we will use an antiferromagnetic state of alternating $\ua$ and $\da$ spins, since this will capture the dynamics of the exchange interaction between adjacent spins. This is a reasonable choice, but a different choice would not change the theory at all, only the quantitative details of the results. Then beginning with the initial state $\ket{\Psi_0}=\ket{\ua\da\ua\da}$ and following a perturbative approach identical to Sec. III A, we obtain the following disorder averaged return probability:

\begin{align*}
&[P(t)]=\frac{7}{18}-\frac{13\sigma_J^2}{24J_0^2}\\
&+\bigg(\frac{1}{3}e^{-\sigma_J^2t^2/8}+\frac{2\sigma_J^2}{3J_0^2}e^{-\sigma_J^2t^2/4}-\frac{5\sigma_J^2}{16J_0^2}e^{-\sigma_J^2t^2/8}\bigg)\cos J_0t\\
&+\bigg(\frac{1}{6}e^{-\sigma_J^2t^2/2}+\frac{\sigma_J^2}{3J_0^2}e^{-5\sigma_J^2t^2/8}-\frac{\sigma_J^2}{8J_0^2}e^{-\sigma_J^2t^2/2}\bigg)\cos 2J_0t\\
&+\bigg(\frac{1}{9}e^{-9\sigma_J^2t^2/8}-\frac{\sigma_J^2}{48J_0^2}e^{-9\sigma_J^2t^2/8}\bigg)\cos 3J_0t+O\bigg(\frac{\sigma_J^3}{J_0^3},\frac{\sigma_j^2t}{J_0}\bigg)
\numb
\label{eqn:pave4q}
\end{align*}

The unperturbed Hamiltoninan has energies of $J_0$, $0$, $-J_0$ and $-2J_0$, the differences of which lead to the three frequencies present in $[P(t)]$. A four qubit linear array breaks some symmetry of the system, and causes degenerate states to split into six different energies, causing many more frequencies to appear.

\section{Discussion}

Using exact analytical methods, we have derived expressions for the disorder averaged return probability from a given initial state for a system consisting of a ring of three exchange coupled qubits. We have also calculated quantities such as the normalized Hamming distance, qubit spin expectation values, and entanglement entropy for each qubit. We then used a perturbative approach to analyze a linear array of three qubits, comparing our results to direct numerics, and extended the same perturbative approach to a ring of four qubits. The frequencies by which the disorder averaged return probability oscillates are determined by the energy differences present in the unperturbed Hamiltonian. These frequencies have a Gaussian-like decay, in contrast to the usually assumed exponential decoherence ansatz, with their decay rate generally dependent on the first-order correction to the corresponding energies. For the system with a 3-qubit ring, the high degree of symmetry in the Hamiltonian protects states of different energies from mixing due to disorder, and so the shape of the oscillation envelope is completely independent of the mean exchange strength $J_0$, instead depending only on the strength of the disorder $\sigma_J$.

We compare our results to the analogous two-qubit calculation in Ref. \onlinecite{ThrockmortonPRB2017}, where the return probability was found to be given by:

\begin{equation}
[P(t)]=\frac{1}{2}+\frac{1}{2}e^{-\sigma_J^2t^2/2}\cos J_0t
\numb
\label{eqn:pave2q}
\end{equation}

Here the oscillation envelope is strictly Gaussian, and we showed that with higher-qubit systems, the envelope keeps its Gaussian-like characteristics, though the shape is no longer exactly Gaussian. In Table \ref{tab:decay}, we calculate the time $t_c$ needed for the width of the envelope to reach a fraction of its initial value, for each system and for various coherence fraction. $T_2^*$ is given by the time required for the envelope width to decay to $1/e$ of its original value (the bottom row in Table \ref{tab:decay}), and the times corresponding to values such as $.99$ are used to benchmark gates with that fidelity. For the 3-qubit line and 4-qubit ring, the envelope shape depends partly on $J_0$, so we give results in the limit that $J_0/\sigma_J\rightarrow\infty$. Because of the Gaussian-like nature of the envelopes, for $t\ll\sigma_J^{-1}$, the envelopes have a parabolic shape, as demonstrated by expansions such as Eq. (\ref{eqn:paveexpand0}). This is evident in Table \ref{tab:decay}, as changing the target decoherence by a factor of 100 changes $t_c$ by roughly a factor of 10. We also see that increasing the number of qubits from two to three decreases the coherence time, but the 4-qubit ring has increased coherence time from the others. It is important to note that the 3-qubit line and 4-qubit ring have multiple frequencies present in $[P(t)]$ which decay at different rates, and thus the coherence times are heavily dependent on the initial state, as a different initial state will lead to different weights for each frequency. The fact that the coherence time decreases by roughly 10\% in going from 2 to 3 qubits is of considerable significance indicating that maintaining coherence in larger qubit systems would be increasingly a challenge unless the system has special symmetries (e.g. 3 and 4 qubit rings) which can be exploited to enhance coherence.  In addition, the fact that decoherence happens through a Gaussian temporal decay and not exponentially is also of substantial significance. We should mention that scaling up number of qubits necessitates careful consideration of how decoherence changes, since a short decoherence time generates larger errors making fault-tolerant quantum gate operations increasingly more challenging.

\acknowledgements

This work is supported by the Laboratory for Physical Sciences.

\end{document}